\documentclass[twocolumn,superscriptaddress,nobalancelastpage,10pt,pra,a4paper]{revtex4}
\usepackage{amsmath}
\usepackage{amsthm}
\usepackage{amssymb}
\usepackage{amsfonts}
\usepackage{graphicx}
\usepackage{wasysym}
\usepackage{mathrsfs}
\usepackage{yfonts}
\usepackage{bbold}
\usepackage{verbatim}
\usepackage{subfigure}
\usepackage{bbm}
\usepackage{color}

\newcommand{\ket}[1]{\left| #1 \right\rangle}
\newcommand{\bra}[1]{\left\langle #1 \right|}

\newcommand{\opa}{\widehat{a}}

\newcommand{\sinc}[1]{\text{sinc}[#1]}

\newcommand{\roml}[1]{\lowercase\expandafter{\romannumeral #1\relax}}
\newcommand{\romu}[1]{\uppercase\expandafter{\romannumeral #1\relax}}

\DeclareMathAlphabet{\mathpzc}{OT1}{pzc}{m}{it}

\begin{document}

\title{Nonclassicality of induced coherence without induced emission}

\author{Mayukh Lahiri}
\email{mlahiri@okstate.edu} \affiliation{Department of Physics, Oklahoma State University,
Stillwater, Oklahoma 74078-3072, USA}

\author{Armin Hochrainer} \affiliation{Vienna Center for
Quantum Science and Technology (VCQ), Faculty of Physics, University
of Vienna, Boltzmanngasse 5, Vienna A-1090,
Austria.}\affiliation{Institute for Quantum Optics and Quantum
Information, Austrian Academy of Sciences, Boltzmanngasse 3, Vienna
A-1090, Austria.}

\author{Radek Lapkiewicz} \affiliation{Institute of Experimental Physics,
Faculty of Physics, University of Warsaw, Pasteura 5, Warsaw 02-093, Poland}

\author{Gabriela Barreto Lemos} \affiliation{Physics Department,
University of Massachusetts Boston, 100 Morrissey Blvd., Boston MA 02125, USA}

\author{Anton Zeilinger}\affiliation{Vienna Center for
Quantum Science and Technology (VCQ), Faculty of Physics, University
of Vienna, Boltzmanngasse 5, Vienna A-1090,
Austria.}\affiliation{Institute for Quantum Optics and Quantum
Information, Austrian Academy of Sciences, Boltzmanngasse 3, Vienna
A-1090, Austria.}
\begin{abstract}
\noindent Interference of two beams produced at separate biphoton
sources was first observed more than two decades ago. The
phenomenon, often called ``induced coherence without induced
emission'', has recently gained attention after its applications to
imaging, spectroscopy, and measuring biphoton correlations have been
discovered. The sources used in the corresponding experiments are
nonlinear crystals pumped by laser light. The use of a laser pump
makes the occurrence of induced (stimulated) emission unavoidable
and the effect of stimulated emission can be observed in the joint
detection rate of the two beams. This fact raises the question
whether the stimulated emission also plays a role in inducing the
coherence. Here we investigate a case in which the crystals are
pumped with a single-photon Fock state. We find that coherence is
induced even though the possibility of stimulated emission is now
fully ruled out. Furthermore, the joint detection rate of the two
beams becomes ideally zero. Our results rule out any classical or
semi-classical explanation of the phenomenon and also suggest that
it is, in principle, possible to perform similar experiments
with fermions, for which stimulated emission is strictly forbidden.
\end{abstract}

\maketitle

\section{Introduction}
\begin{figure}[b]
    \centering
        \includegraphics[width=0.45\textwidth]{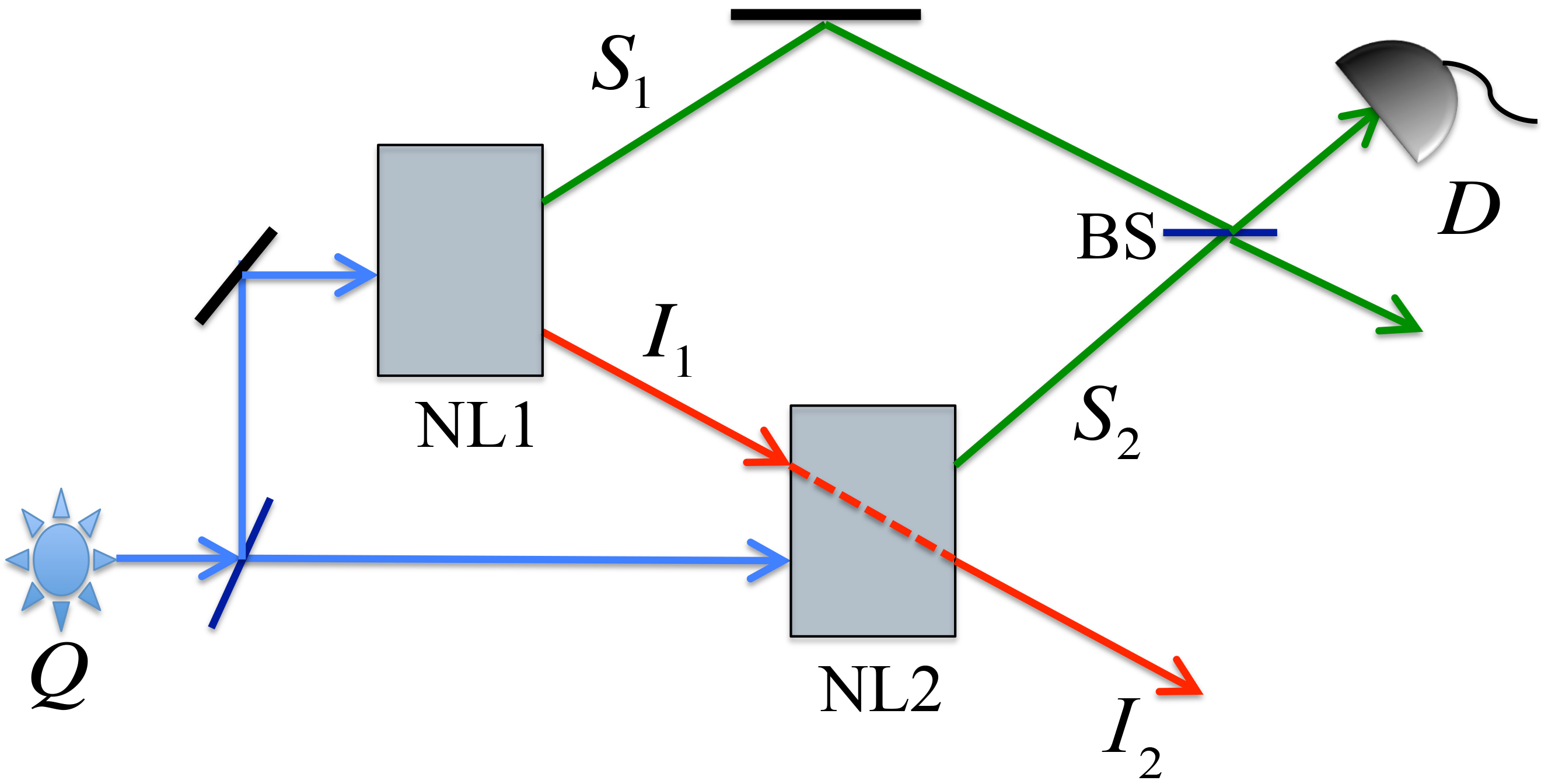}
\caption{Schematic of the experimental setup. $Q$ is a source that
generates the pump beam. In our proposed experiment it is a
single-photon source. (In usual experiments it is a laser source.)
The pump beam is split into two to illuminate two identical
nonlinear crystals, NL1 and NL2. The crystals generate signal
($S_1$, $S_2$) and idler ($I_1$, $I_2$) beams by the process of
parametric down-conversion. Signal beams $S_1$ and $S_2$ are
combined by a beam splitter $BS$ and the resulting beam is collected
by a detector D. The idler beams are aligned through NL2 but never
detected.}
    \label{fig-ZWM}
\end{figure}

In 1991 \cite{ZWM-ind-coh-PRL,WZM-ind-coh-PRA}, Zou, Wang, and
Mandel (ZWM) induced coherence between two light beams generated by
two spatially separated identical biphoton sources. The sources were
nonlinear crystals each of which could emit two photons into two
separable beams. The crucial technique (originally suggested by Ou)
was to send a beam from one of the sources through the other source
and to overlap it with the beam of the same photon generated by the
latter (Fig. \ref{fig-ZWM}). In a recent series of work, this
technique has been used for imaging
\cite{LBCRLZ-mandel-im,LLLZ-th-mandel-img}, spectroscopy
\cite{Kulik-spec-2016}, generating a light beam in any state of
polarization \cite{LHLBZ-quant-pol}, testing the complementarity
principle \cite{HMM-comp-two-ph-PRL,HMM-comp-two-ph-PRA}, two-color
interferometry \cite{HLLLZ-eq-wvln-Optica}, measuring correlations
between two photons
\cite{HLLLZ-mom-corr-exp-PNAS,LHLLZ-mom-corr-th-PRA}, and generating
many-particle entangled states
\cite{ent-path-id-PRL-2017,L-MPPI-PRA-2018}.
\par
In all the above-mentioned experiments, laser light has been used to
pump the identical nonlinear crystals. Therefore, when a beam from
one of the sources is sent through the other source, the occurrence
of stimulated (induced) emission becomes, in principle, unavoidable.
This fact leads to the question whether stimulated emission is the
cause of the induced coherence.
This question is important to address mainly because of two reasons.
1) It would otherwise leave the possibility of explaining the
generation of coherence semi-classically. 2) More importantly, if
stimulated emission plays a key role in inducing the coherence, it
will not be possible to perform similar experiments with fermions
for which stimulated emission is strictly forbidden.
\par
Zou, Wang, and Mandel \cite{ZWM-ind-coh-PRL,WZM-ind-coh-PRA}
claimed that stimulated emission is not the cause of induced
coherence. This point has been reinvestigated later by Wiseman and
M{\o}lmer \cite{WM-ind-coh-cl-q} and, again recently, by Shapiro et
al. \cite{shapiro-criticism}. However, the conclusion of Refs.
\cite{ZWM-ind-coh-PRL,WZM-ind-coh-PRA,WM-ind-coh-cl-q} is in direct
contradiction to the conclusion of Ref. \cite{shapiro-criticism}:
while Refs. \cite{ZWM-ind-coh-PRL,WZM-ind-coh-PRA,WM-ind-coh-cl-q}
conclude that stimulated emission does not play any role in inducing
coherence, Ref. \cite{shapiro-criticism} claims exactly the opposite
(further details are given in Sec. \ref{sec:discuss}).
\par
We address this question from a new angle. We show that if
the biphoton sources are pumped with a single-photon Fock state,
stimulated emission becomes forbidden but the two beams remain
mutually coherent. Our results thus establish that stimulated
emission is not responsible for the mutual coherence and, therefore,
suggest that performing such experiments with fermions is, in
principle, possible. We also discuss the similarities and
differences between the cases of a single-photon pump and a laser
pump.

\section{Description of The Proposed Experiment}
A ZWM-interferometer (Fig. \ref{fig-ZWM}) uses two spatially
separated identical biphoton sources (NL1 and NL2). Each source
can emit a photon pair (signal and idler) into a pair of
beams. NL1 emits signal and idler photons into beams $S_1$ and $I_1$
respectively. Similarly, NL2 emits the photons into $S_2$ and $I_2$.
In the experiment, single-photon interference between $S_1$ and
$S_2$ is observed by erasing which-path information with the help of
$I_1$ and $I_2$. The key is to send $I_1$ through NL2 and to align
it with $I_2$. In such experiments, weakly pumped nonlinear crystals
are used as biphoton sources. Since these crystals are usually
pumped by \emph{laser beams}, there is always a non-zero probability
of the presence of idler photons generated by the first source at
the second source when the down-conversion is taking place at the
latter. Therefore, occurrence of stimulated (induced) emission at
NL2 becomes, in principle, inevitable.
\par
We propose to replace the pumping laser source with a
\emph{single-photon source}. The light generated by the
single-photon source displays antibunching \cite{KDM-antibunch}.
Below we discuss in detail the expected outcome of the experiment.

\section{Theory}
In the process of parametric down-conversion, a nonlinear crystal
absorbs a pump photon and generates a photon pair (signal and
idler); the generation of multiple pairs is also possible depending
on the state of the pumping field. For simplicity, we consider
single-mode optical fields. We denote the frequencies of pump,
signal, and idler fields by $\omega_P$, $\omega_S$, and $\omega_I$,
respectively. The Hamiltonian describing parametric down-conversion
at either crystal can be expressed, in the interaction picture, as
(see, for example, \cite{GHOM-PRA})
\begin{equation}\label{Ham-j}
\widehat{H}_j(t)=g'e^{i\Delta\omega t}\opa_{Pj}
\opa^{\dag}_{Sj}\opa^{\dag}_{Ij}+\text{H.c.},
\end{equation}
where $j=1,2$ labels the two nonlinear crystals, $g'$ represents the
interaction strength; $P$, $S$, and $I$ represent pump, signal, and
idler photons respectively; $\opa$ and $\opa^{\dag}$ represent
photon annihilation and creation operators respectively;
$\Delta\omega=\omega_S+\omega_I-\omega_P$; and H. c. denotes
Hermitian conjugation.  The quantum state of light generated by each
crystal is obtained by the standard perturbative solution (see, for
example, \cite{Cohen-Tannoudji-QM}) and is given by the well-known
expression
\begin{align}\label{state-s-1-2-gen}
&\ket{\psi_j}= \Big[\mathbbm{1}+\frac{1}{i\hbar}\int_0^{\tau}dt_1
\widehat{H}_j(t_1) \nonumber \\
&+\left(\frac{1}{i\hbar}\right)^2\int_0^{\tau}dt_1
\widehat{H}_j(t_1)\int_0^{t_1}dt_2
\widehat{H}_j(t_2)+\dots\Big]\ket{\psi_{j0}} \nonumber \\
&\equiv \widehat{\mathscr{U}_j}\ket{\psi_{j0}},
\end{align}
where $\ket{\psi_{j0}}$ is the state of light before down
conversion, $\tau$ is the interaction time which is usually the time
taken by the pump to travel the crystal's length, and we have
dropped the normalization constant. It is important to note that the
interaction Hamiltonian is time dependent and one needs to consider
the proper ordering of the time-integrations while calculating the
higher order terms.
\par
For example, by carrying out the integrations in Eq.
(\ref{state-s-1-2-gen}) and dropping the terms with zero
contribution, we can express the state generated by NL1 in the
following form:
\begin{align}\label{state-s-1-2}
\ket{\psi_1}&= \Big[\mathbbm{1}+g\opa_{P1}\opa^{\dag}_{S1}
\opa^{\dag}_{I1}
+\frac{g^2}{2}\left(\opa_{P1}\opa^{\dag}_{S1}\opa^{\dag}_{I1}\right)^2
\nonumber \\ & \qquad +\tilde{g}^2\opa^{\dag}_{P1}\opa_{P1}\opa_{S1}
\opa^{\dag}_{S1}\opa_{I1}\opa^{\dag}_{I1}+\dots
\Big]\ket{\psi_{10}},
\end{align}
where $g$ and $\tilde{g}$ contains the same order of $g'$. Although
their explicit forms are not necessary for the purpose of our
discussion, they are given in Appendix.
\par
If the $I_1$ beam originating from NL1 is sent through NL2 and then
perfectly aligned with the $I_2$ beam (Fig. \ref{fig-ZWM}), we have
\begin{align}\label{align-cond}
\opa_{I_2}=\opa_{I_1}\exp[i\phi_I],
\end{align}
where $\phi_I$ is the phase change due to propagation from crystal
NL1 to crystal NL2.
\par
When the two crystals are put into the ZWM setup (Fig.
\ref{fig-ZWM}), the quantum state of light generated by them is
given by (cf. \cite{LLLZ-th-mandel-img,LSGHOG-ind-emiss})
\begin{equation}\label{state-tot-form-2}
\ket{\Psi}=\widehat{\mathscr{U}_2}\widehat{\mathscr{U}_1}
\ket{\psi_0},
\end{equation}
where $\ket{\psi_0}$ is the initial state of light before any
down-conversion took place, and Eq. (\ref{align-cond}) has been
substituted into the expression of $\widehat{\mathscr{U}_2}$.
Equation (\ref{state-tot-form-2}) is applicable to both the cases
where laser and single-photon pumps are used. It is the initial
state, $\ket{\psi_0}$, which makes the difference between the two
cases.
\par
We stress that this theoretical treatment is valid only when
the interaction is sufficiently weak.

\subsection{Pumping with Single-photons}
\par
A single-photon source produces light that displays antibunching
\cite{KDM-antibunch}, i.e. if the light is sent through a beam
splitter, no coincidence counts between the two outputs are
registered for an appropriate choice of the delay line. This is
because a single photon cannot be broken into further halves and,
therefore, cannot be absorbed at more than one place simultaneously.
In the experiment (Fig. \ref{fig-ZWM}), the two crystals are placed
at the two outputs of a beam splitter and, therefore, they cannot
both absorb the same pump photon. Consequently, if the time
difference between two consecutive pump photons is sufficiently
large, it becomes impossible for an idler photon generated by NL1 to
be present at NL2 when down-conversion takes place in the latter.
The occurrence of stimulated emission at NL2 thus becomes strictly
forbidden. We prove below that even in this case the signal beams,
$S_1$ and $S_2$, remain fully coherent.
\par
We assume, for simplicity, that the pump beam has the same intensity
at the two crystals. When the pump beam is generated by a
single-photon source, the initial state [see Eq.
(\ref{state-tot-form-2})] is given by
\begin{equation}\label{state-in-1}
\ket{\psi_0}_{sp}=\frac{1}{\sqrt{2}}\left(\ket{1_{P_1}}+e^{i\phi_P}
\ket{1_{P_2}}\right) \ket{\text{vac}_{\{S,I\}}},
\end{equation}
where ``vac'' implies vacuum (no photon),
e.g.,$\ket{\text{vac}_{\{S,I\}}}$ signifies no occupation in the
signal and the idler modes generated by both crystals; $\phi_P$
represents the phase difference between the pump field at the two
crystals; and the single-photon Fock state $\ket{1_{P_j}}$
represents a pump photon for the crystal $j$ such that
$\opa_{P_j}\ket{1_{P_j}}=\ket{\text{vac}_{P_j}}$. It now follows
from Eqs. (\ref{state-s-1-2-gen})--(\ref{state-in-1}) that the
quantum state of light in the system takes the form
\begin{align}\label{state-tot-no-ob-1}
\ket{\Psi_{sp}}=& \eta\ket{\psi_0}_{sp} \nonumber \\ &
+G_{sp}\ket{\text{vac}_{\{P\}}}\Big[\left(\ket{1_{S_1}}
+e^{i\phi}\ket{1_{S_2}}\right) \ket{1_{I_1}}\Big],
\end{align}
where $\eta$ has contributions from all even orders of $g'$,
$G_{sp}$ has contributions from all odd orders of $g'$; $|\eta|\gg
|G_{sp}|$ and $\phi=\phi_P-\phi_I$. It is important to note that Eq.
(\ref{state-tot-no-ob-1}) provides an exact expression that is
obtained without dropping any higher-order term; the terms
containing even order of $g'$ in
$\widehat{\mathscr{U}_2}\widehat{\mathscr{U}_1}$ yield the initial
state $\ket{\psi_0}_{sp}$, and the terms containing odd order of
$g'$ yield the state multiplied with $G_{sp}$.
\par
In order to calculate the photon counting rate, we need to determine
the quantized electric field at the detector, $D$ (placed after the
beam splitter, $BS$, in Fig. \ref{fig-ZWM}). The positive frequency
part of the field can be represented by
\begin{align}\label{quant-E-det}
\widehat{E}_S^{(+)}=\widehat{a}_{S_1}+ie^{i\phi_S}\widehat{a}_{S_2},
\end{align}
where $\phi_S$ is the phase due to the difference between the
optical paths from $NL1$ and $NL2$ to $D$. The photon counting rate
at $D$ is then given by
\begin{align}\label{R-at-det-no-ob-1}
\mathcal{R}_{sp}=\bra{\Psi_{sp}}\widehat{E}_S^{(-)}\widehat{E}_S^{(+)}\ket{\Psi_{sp}}
=2|G_{sp}|^2(1+\cos\phi_{in}),
\end{align}
where $\widehat{E}_S^{(-)}=\{\widehat{E}_S^{(+)}\}^{\dag}$,
$\phi_{in}=\phi_S+\phi+\pi/2$, the state $\ket{\Psi_{sp}}$ is given
by Eq. (\ref{state-tot-no-ob-1}), and we have dropped a constant
multiplicative factor that depends on the detection efficiency. We
note that Eq. (\ref{R-at-det-no-ob-1}) gives an exact expression for
the photon counting rate, i.e. \emph{no higher order term has been
neglected}.
\par
It is clear from Eq. (\ref{R-at-det-no-ob-1}) that the signal beams
$S_1$ and $S_2$ create a single-photon interference pattern at the
detector $D$. This means that the two signal beams are mutually
coherent. Since the occurrence of stimulated emission is forbidden
(see discussions above), this mutual coherence in the lowest-order
\cite{Note-corr-order} can only be explained from the
indistinguishability of the paths for the signal photons arriving at
the detector.
\par
\begin{figure}
    \centering
        \includegraphics[width=0.45\textwidth]{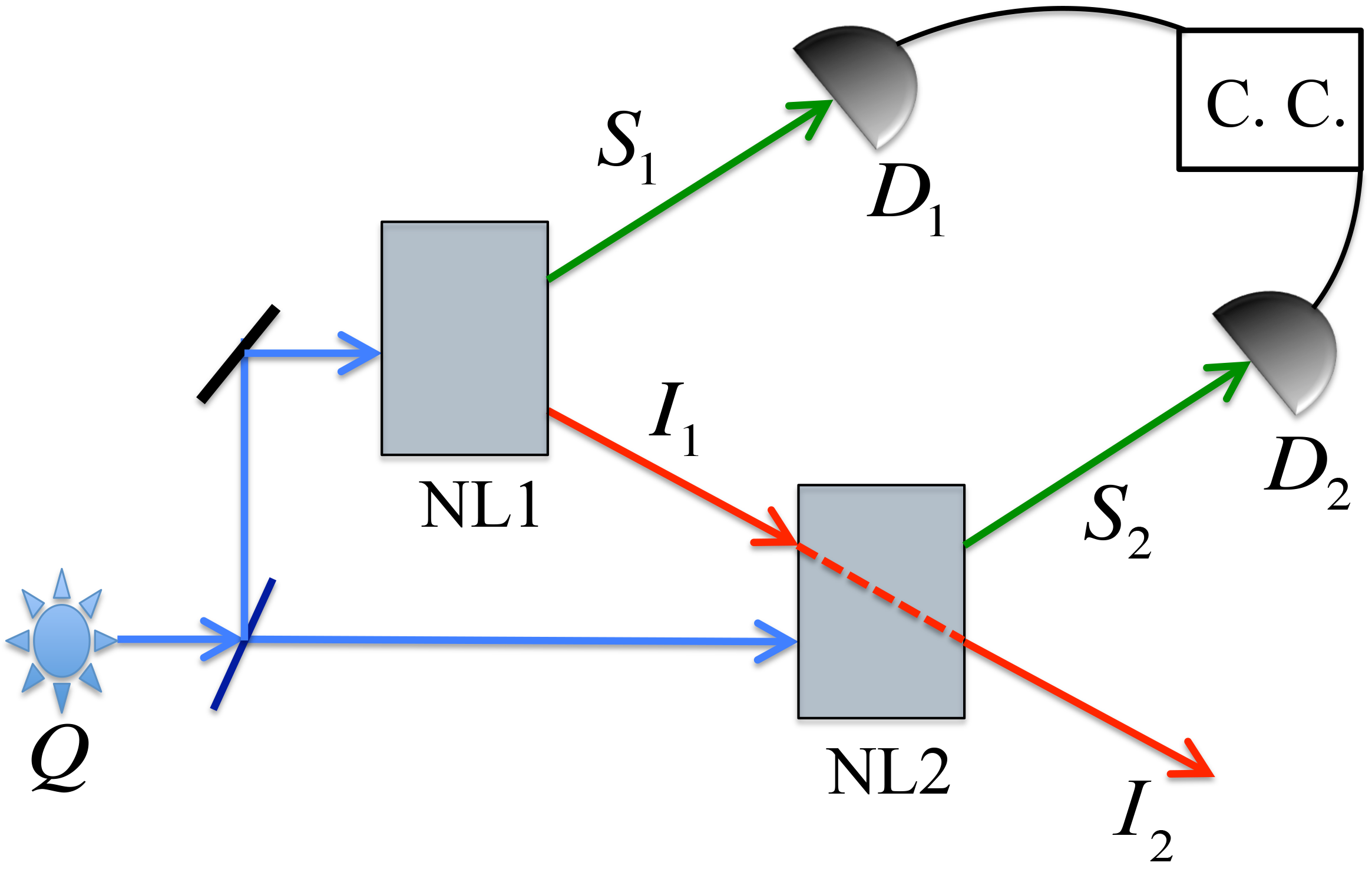}
\caption{\small{Setup for detecting the effect of stimulated
emission in coincidence counting. Coincidence detections between
beams $S_1$ and $S_2$ are registered with detectors $D_1$ and $D_2$.
The optical path length from $Q$ to $D_1$ via $NL1$ is equal to the
optical path length from $Q$ to $D_2$ via $NL2$.}}
    \label{figb}
\end{figure}

The single-photon pump ensures that only one pair of down-converted
photons exists in the system. This fact is also justified by the
absence of terms with higher photon number in Eq.
(\ref{state-tot-no-ob-1}). Since the signal modes generated by both
crystals are never simultaneously occupied, there will be no
coincidence count if one detects both signal beams (Fig.
\ref{figb}). This point is elaborated in the next section.

\subsection{Comparison with the Case of Laser Pump}
\par
We represent the laser field by a coherent state. When the two
crystals are pumped by a laser, the initial state (before any
down-conversion) is given by
\begin{equation}\label{state-in-2}
\ket{\psi_0}_{lp}=\ket{\alpha_1}_P\ket{\alpha_2}_P
\ket{\text{vac}_{\{S,I\}}},
\end{equation}
where the suffix, $lp$, denotes laser pump,
$\ket{\text{vac}_{\{S,I\}}}$ signifies zero occupation in all signal
and idler modes, and $\ket{\alpha_j}_P$ represents the coherent
state of the pump at crystal $j$ such that
$\opa_{P_j}\ket{\alpha_j}_P=\alpha_j\ket{\alpha_j}_P$. For
simplicity, we again assume that the pump beams have the same
intensities at the two crystals, i.e.
$|\alpha_1|=|\alpha_2|=\alpha$, say. It now follows from Eqs.
(\ref{state-tot-form-2}) and (\ref{state-in-2}) that
\begin{align}\label{state-tot-no-ob-2}
\ket{\Psi_{lp}} & \approx \ket{\psi_0}_{lp}+\bigg[
g_{lp}\alpha\ket{\{\alpha_P\}} \left(\ket{1_{S_1}}
+e^{i\phi}\ket{1_{S_2}}\right) \ket{1_{I_1}}\bigg] \nonumber
\\
& +\bigg[(g_{lp}\alpha)^2\ket{\{\alpha_P\}}
\Big(\ket{2_{S_1}}+e^{2i\phi}\ket{2_{S_2}}\nonumber \\ & \qquad
\qquad \qquad \qquad
+\sqrt{2}e^{i\phi}\ket{1_{S_1},1_{S_2}}\Big)\ket{2_{I_1}} \nonumber
\\
& \qquad +
\tilde{g}_{lp}^2\left(\alpha_1\opa^{\dag}_{P_1}+\alpha_2\opa^{\dag}_{P_2}
\right)\ket{\psi_0} \bigg]+\dots,
\end{align}
where we have written $g$ of Eq. (\ref{state-s-1-2}) as $g_{lp}$ to
distinguish the case of a laser pump and absorbed the phase factor
$\exp[i\text{arg}\{\alpha_1\}]$ into it;
$\ket{\{\alpha_P\}}=\ket{\alpha_1}_P\ket{\alpha_2}_P$; the terms
containing the same order of $g'$ are arranged inside the same
square brackets; $\phi=\phi_P-\phi_I$ with
$\phi_P=\text{arg}\{\alpha_2\}-\text{arg}\{\alpha_1\}$; and we have
dropped the normalization constant.
\par
We now determine the photon counting rate at $D$ under the
assumptions that the crystals are weakly pumped and the rate of
down-conversion is low, i.e. $|g_{lp}\alpha|^2\ll 1$. It follows
from Eqs. (\ref{quant-E-det}) and (\ref{state-tot-no-ob-2}) that the
photon counting rate is given by
\begin{align}\label{R-at-det-no-ob-2}
\mathcal{R}_{lp}=\bra{\Psi_{lp}}\widehat{E}_S^{(-)}\widehat{E}_S^{(+)}\ket{\Psi_{lp}}
\approx 2|{g}_{lp}\alpha|^2(1+\cos\phi_{in}),
\end{align}
where $\phi_{in}=\phi_S+\phi+\pi/2$ and we have dropped a constant
multiplicative factor that depends on the detectors's efficiency.
We find that Eq. (\ref{R-at-det-no-ob-2}) and Eq.
(\ref{R-at-det-no-ob-1}) have the same form. A comparison between
the cases of single-photon pump and laser pump reveals many
interesting features as we discuss below.
\par
We note that for a single-photon pump, the quantum state [Eq.
(\ref{state-tot-no-ob-1})] does not contain any term that involves
more than two down-converted photons. In contrast, the quantum state
for the case of a laser pump [Eq. (\ref{state-tot-no-ob-2})]
contains terms that involve more than two down-converted photons. In
the single-mode treatment, these terms correspond to stimulated
emission \cite{Note-sp-em}. Furthermore, these terms are also
multiplied by the second or higher powers of $g_{lp}\alpha$.
Neglecting $|g_{lp}\alpha|^2$ with respect to 1 is, therefore,
equivalent to neglecting the effect of stimulated emission. We find
from Eq. (\ref{R-at-det-no-ob-2}) that this approximation does
\emph{not} destroy the interference pattern. Furthermore, the
visibility of the patterns for both types of pumps are equal. We
thus conclude that \emph{although stimulated emission occurs when a
laser pump is used, the mutual coherence between the two signal
beams is not due to stimulated emission}; the spontaneous emissions
occurring at the two crystals play the dominating role when the
crystals are weakly pumped and the rate of down conversion is low.
\par
A close examination of Eq. (\ref{state-tot-no-ob-2}) reveals
that two kinds of stimulated emission are present for a laser pump:
\roml{1}) the terms containing $\ket{2_{S_1}}$ and $\ket{2_{S_2}}$
correspond to the emission stimulated by photons generated in the
same crystal; and \roml{2}) the term containing
$\sqrt{2}\ket{1_{S_1},1_{S_2}}$ refers to the emission at NL2
stimulated by the idler photon generated at NL1. We also note that
if the beam $I_1$ is blocked between NL1 and NL2, the term
$\sqrt{2}\ket{1_{S_1},1_{S_2}}\ket{2_{I_1}}$ gets replaced by
$\ket{1_{S_1},1_{I_1},1_{S_2},1_{I_2}}$. Therefore, the alignment of
the idler beams enhances the probability of joint pair emission at
NL1 and NL2 by a factor of two. This effect can be observed in the
intensity correlation of the signal beams.
\begin{figure}[htbp]
    \centering
        \includegraphics[width=0.45\textwidth]{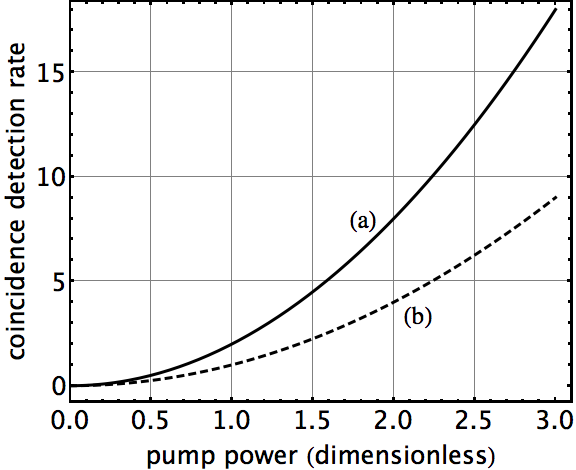}
\caption{Effect of stimulated emission in the case of a laser
pump. The coincidence detection rate (dimensionless unit) of two
signal photons ($S_1$ and $S_2$) is plotted against the pump power
($|\alpha|^2$; dimensionless unit) for the two cases: (a) the idler
beams are fully aligned; and (b) the idler beams are fully
misaligned. In both cases, the coincidence detection rate increases
quadratically with the pump power. However, if the idlers are fully
misaligned, the coincidence rate becomes two times smaller than the
value obtained for fully aligned idler beams. The stimulated
emission caused by the alignment of idler beams results in the
higher coincidence rate on curve (a) compared to curve (b).}
    \label{fig-las-sp-comp}
\end{figure}
\par
Let us, for example, analyze the situation illustrated in Fig.
\ref{figb}. Here, coincidence detection of $S_1$ and $S_2$ is
considered when any time delay between them is fully compensated by
a coincidence circuit. It follows from Eq. (\ref{state-tot-no-ob-2})
that for a laser pump ($lp$), the coincidence detection rate is
given by
\begin{align}\label{coincidence-las}
\mathscr{C}_{lp}^{S_1,S_2}
=\bra{\Psi_{lp}}\widehat{a}^{\dag}_{S_1}\widehat{a}^{\dag}_{S_2}
\widehat{a}_{S_2}\widehat{a}_{S_1}\ket{\Psi_{lp}}\approx
2|g_{lp}\alpha|^4.
\end{align}
Clearly, the rate of coincidence detection increases
quadratically with the pump power ($|\alpha|^2$). This coincidence
detection rate is due to the joint pair emission at NL1 and NL2,
i.e. due to the term $\sqrt{2}\ket{1_{S_1},1_{S_2}}\ket{2_{I_1}}$ in
Eq. (\ref{state-tot-no-ob-2}). If the idler beams are misaligned,
one has $\sqrt{2}\ket{1_{S_1},1_{S_2}}\ket{2_{I_1}}\rightarrow
\ket{1_{S_1},1_{I_1},1_{S_2},1_{I_2}}$ and the corresponding
coincidence detection rate becomes
\begin{align}
\mathscr{C}_{lp}^{S_1,S_2} \approx |g_{lp}\alpha|^4 \quad
(\text{idlers not aligned}),\label{coincidence-las-na}
\end{align}
Comparing with Eq. (\ref{coincidence-las}) with
(\ref{coincidence-las-na}), we find that a complete misalignment of
the idler beams reduces the $S_1$--$S_2$ coincidence detection rate
by a factor of two (Fig. \ref{fig-las-sp-comp}). The phenomenon can
be physically understood as follows: The presence of a photon in the
$I_1$ mode at NL2, makes the emission of another photon in the same
mode more probable. Since emission of an idler photon is always
accompanied by the emission of its partner signal photon, the
$S_1$--$S_2$ coincidence detection rate enhances.
\par
For the case of a single-photon pump ($sp$), the coincidence rate is
\begin{align}\label{coincidence-sp}
\mathscr{C}_{sp}^{S_1,S_2}
=\bra{\Psi_{sp}}\widehat{a}^{\dag}_{S_1}\widehat{a}^{\dag}_{S_2}
\widehat{a}_{S_2}\widehat{a}_{S_1}\ket{\Psi_{sp}}=0,
\end{align}
where $\ket{\Psi_{sp}}$ is given by Eq. (\ref{state-tot-no-ob-1}).
Clearly, the rate of coincidence detection does not depend on
the crystal gain, i.e. the rate of photon production. In the case of
a single-photon pump, no stimulated emission is possible even with
fully aligned idler beams. Therefore, the $S_1$--$S_2$ coincidence
detection rate ($\mathscr{C}_{sp}^{S_1,S_2}$) remains unchanged
(zero) when the idler beams are misaligned. However, for both types
of pumps, the lowest-order correlation between the two signal beams
will be completely lost if the idler beams are fully misaligned.

\section{Connection with Statistical Properties of Light}\label{sec:ph-stat}
When a nonlinear crystal is pumped with a single-photon Fock state,
the down-converted light is ideally in a two-photon Fock state. This
down-converted light displays different photon statistics than the
down-converted light generated from a laser pump. An undepleted
laser pump, which is usually modeled by a coherent state, allows one
to treat the pump field classically. In this case, signal and idler
beams are individually in thermal (chaotic) states
\cite{MG-pdc-PR-2,2004-PA-pdc-therm-stat-exp}; the quantum state
produced by the crystal is a superposition of Fock states, each of
which contains equal number of signal and idler photons. However, a
single-photon pump cannot be treated classically; in this case, each
down-converted beam is individually represented by a single-photon
Fock state which is certainly not a thermal state. This difference
between the statistical properties of the down-converted light can
be intuitively connected to our results.
\par
It is well-known that the phase space distributions of Fock states
are not Gaussian (\cite{MW}, Sec. 11.8.6) and, consequently, the
higher-order field correlation functions cannot, in general, be
expressed in terms of the lowest-order ones \cite{G-ph-corr}. The
analogous result in our case is the fact that when $S_1$ and $S_2$
beams contain photons in a Fock state, the lowest-order coherence
between the beams is \emph{not} accompanied by any intensity
correlation.
\par
The phase space distribution of light in a thermal (chaotic) state
is Gaussian (see, for example, \cite{MW}, Sec. 11.8.6). For such
light, all higher-order field correlation functions can be expressed
in terms of the lowest-order ones (\cite{MW}, Ch. 13). We have found
in our analysis that in the case of thermal $S_1$ and $S_2$ beams
(produced by laser pumps), the lowest-order coherence
\cite{Note-corr-order} is necessarily accompanied by the presence of
the intensity correlation between the two beams. When the idler
beams are misaligned, the loss of lowest-order coherence between
$S_1$ and $S_2$ is associated with the reduction of the coincidence
detection rate of the two beams. As discussed above this reduction
is due to the fact that photons generated at NL2 are not stimulated
by the emission at NL1.
\par
One can therefore conclude that for a laser pump, the induced
lowest-order coherence needs to be accompanied by stimulated
emission, even though stimulated emission is not responsible for
inducing coherence.

\section{Existing Experimental Evidence}\label{sec:exp-res}
With the available technology, it is extremely challenging to
perform the above mentioned experiments with a single-photon pump
\cite{Note-Boyd}. There is an alternative way of showing that
stimulated emission at NL2 plays no role in inducing lowest-order
coherence between the two signal beams. This method, which was
introduced in Ref. \cite{ZWM-ind-coh-PRL}, is to insert an
attenuator on the path of the idler beam between NL1 and NL2 and
then to show that the visibility of the interference pattern is
linearly proportional to the amplitude transmission coefficient of
the attenuator. It was later shown in Ref. \cite{WM-ind-coh-cl-q}
that this dependence does not remain linear when the effect of
stimulated emission is prominent. A recent paper also analyzes this
issue in great detail \cite{Kolbov-Boyd-jopt-2017}.
\par
The coincidence measurement between $S_1$ and $S_2$ for laser pump
has been performed by Liu, et al., under more general
considerations, where they controlled the rate of stimulated
emission by placing an attenuator in the idler's path between NL1
and NL2 \cite{LSGHOG-ind-emiss}. They observed a significant drop in
the coincidence detection rate when the idler beam was fully blocked
(no induced emission), compared to the case when the idler beam was
fully transmitted (maximum induced emission).

\section{Discussions}\label{sec:discuss}
\par
In order to put our results into the context of existing work,
we now briefly discuss the arguments presented in Refs.
\cite{ZWM-ind-coh-PRL,WZM-ind-coh-PRA,WM-ind-coh-cl-q,shapiro-criticism}.
\par
Zou, Wang, and Mandel stated that ``when the external field is
weak, the down-conversion occurs spontaneously and at random''
\cite{WZM-ind-coh-PRA}. Here, the external field corresponds to beam
$I_1$ and the down-conversion refers to the down-conversion at NL2.
A weak field implies that the average photon number generated by
down-conversion is very low. Their theoretical analysis, based on
this assumption, correctly predicted the experimentally observed
\cite{ZWM-ind-coh-PRL,WZM-ind-coh-PRA} linear dependence of
visibility on the amplitude transmission coefficient of the
attenuator (see Sec. \ref{sec:exp-res}).
\par
In Ref. \cite{WM-ind-coh-cl-q}, Wiseman and M{\o}lmer tested
whether the above-mentioned linear dependence can be found when
stimulated emission is the cause of induced coherence. They
considered a situation in which the emission rates of the crystals
can be arbitrarily enhanced. They determined the modulus of the
\emph{normalized} lowest-order coherence function between the fields
of the two superposed beams ($S_1$ and $S_2$) and expressed it as a
function of the average photon number generated by down-conversion
(\cite{WM-ind-coh-cl-q}, Eq. (11)). This normalized coherence
function gives the highest attainable visibility. They found that if
the average photon number is high enough for the stimulated emission
to be the cause of induced coherence, the visibility is \emph{not}
proportional to the amplitude transmission coefficient of the
attenuator. However, when the mean photon number is very low, i.e.
when the photon generation is dominated by spontaneous emission, the
visibility becomes proportional to the amplitude transmission
coefficient. Based on this observation they confirmed that the
linear dependence of the visibility on the amplitude transmission
coefficient is the true signature of induced coherence without
stimulated emission.
\par
In Ref. \cite{shapiro-criticism}, Shapiro, Venkatraman, and
Wong considered a situation that is similar to the situation
considered in Ref. \cite{WM-ind-coh-cl-q}, albeit they used a
different terminology. In contrast to Ref. \cite{WM-ind-coh-cl-q},
the lowest-order coherence function presented in Eq. (22) of Ref.
\cite{shapiro-criticism} is not normalized. This \emph{unnormalized}
coherence function decreases with the average photon number and can
give the impression that stimulated emission is responsible for the
induced coherence. It is important to note that the absolute value
of this unnormalized coherence function would always decrease with
decreasing average photon number, even though its normalized version (the
highest attainable visibility) could show an entirely different
behavior. It can be readily checked that if one normalizes the
coherence function obtained by Shapiro et al.
(\cite{shapiro-criticism}, Eq. (22)), one finds the formula derived
by Wiseman and M{\o}lmer (\cite{WM-ind-coh-cl-q}, Eq. (11)). As
mentioned above, this normalized coherence function does not vanish
as the average photon number becomes very low \cite{Note-ph-no-0};
it rather becomes equal to the amplitude transmission coefficient of
the attenuator.
\par
We therefore conclude that in a ZWM-type experiment, the induced
lowest-order coherence is not due to stimulated emission. This fact
rules out a classical or semi-classical interpretation of the
phenomenon like the one suggested in Ref. \cite{shapiro-criticism}.

\section{Summary}\label{sec:summary}
We have proposed to perform a Zou-Wang-Mandel (ZWM) experiment using
a \emph{single-photon pump}. Our theoretical analysis shows that for
such a pump no emission stimulated by the light from the first
source can occur at the second source. We have explicitly shown that
the absence of this stimulated emission does not affect the induced
lowest-order coherence \cite{Note-corr-order} of the two signal
beams, i.e. the beams will produce a single-photon interference
pattern if superposed.
\par
A comparison with the case of laser pump and the existing
experimental evidences establishes that in any ZWM-type experiment,
where the crystals are weakly pumped, the induced lowest-order
coherence is not due to stimulated emission. Therefore, our
results support the conclusion of Refs.
\cite{ZWM-ind-coh-PRL,WZM-ind-coh-PRA,WM-ind-coh-cl-q} and
contradict the conclusion of Ref. \cite{shapiro-criticism}.

\section{Outlook}
\par
As mentioned in Introduction, the ZWM experiment with photons has
found broad applications in quantum optics. Recently, a ZWM-type
experiment experiment has been performed with microwave
superconducting cavities \cite{LPHH-mandel-MSC}. At this point it is
important to look beyond the photonic domain and ask whether similar
experiments can be performed with other quantum entities. Recent
advancements in the fields of trapped ions
\cite{Blatt-Wineland-Nat-2008}, atomic systems
\cite{Keller-Z-PRA-2014,Aspect-at-HOM-2015}, and superconducting
circuits \cite{B-etal-Nat-2014} show very high prospect of research
in this direction.
\par
The ZWM-type experiments performed in the photonic domain suggest
that there is no fundamental obstacle in performing this type of
experiments with other kinds of bosons. However, the case of
fermions require separate attention because stimulated emission is
strictly forbidden in this case.
\par
We have shown that stimulated emission plays no role in the quantum
interference observed in a ZWM experiment. Therefore, our
results suggest that such experiments can, in principle, be
performed with fermions and open up the possibility of building an
experimental setup for this purpose.
\par
Let us briefly illustrate how the same situation can be conceptually
attained with fermions. Consider, for example, electron-electron
scattering. Imagine that electrons are produced in a coherent
superposition of two paths that are incident onto two scattering
samples. One now needs to select only those modes where the
scattered electron paths overlap as in the case of the ZWM
experiment.
\par
Although building such a setup could be technically challenging at
this point, we note that analogs of several optical experiments have
already been performed with fermions. For example, electron
interference has been observed using a Mach-Zehnder setup
\cite{Shtrikman-e-interf}; also, two-particle interference
displaying antibunching (Pauli dip) has been realized with electrons
\cite{Feve-elec-2-pat-interf}. Furthermore, correlated fermions
pairs have also been created in the laboratory \cite{Jin-ferm-pair}.
Based on these facts, we expect that ZWM experiments with fermions
will be performed in the near future.

\section*{Acknowledgments}\label{sec:ackn}
We thank E. Giese for helpful discussions. This work was supported
by the Austrian Academy of Sciences (\"OAW- IQOQI, Vienna), and the
Austrian Science Fund (FWF) with SFB F40 (FOQUS) and W1210-2
(CoQus). R.L. was supported by National Science Centre (Poland)
grants 2015/17/D/ST2/03471, 2015/16/S/ST2/00424, the Polish Ministry
of Science and Higher Education, and the Foundation for Polish
Science (FNP).

\section*{Appendix}
\par
Explicit forms of $g$ and $\tilde{g}$ are obtained by carrying out
the integrations in Eq. (\ref{state-s-1-2-gen}). They are given
below:
\begin{subequations}\label{g-forms-appen}
\begin{align}
& g=\left(\frac{\tau g'}{i\hbar}\right)e^{i\Delta\omega\tau/2}
\sinc{\Delta\omega\tau/2}, \label{g-forms-appen:a}
\\ & \tilde{g}^2=\left(\frac{|g'|}{i\hbar}\right)^2
\frac{i\tau}{\Delta\omega}\left[1+e^{-i\Delta\omega\tau/2}
\sinc{\Delta\omega\tau/2}\right]. \label{g-forms-appen:b}
\end{align}
\end{subequations}


\begin{thebibliography}{99}
\bibitem {Note-corr-order} In coherence theory, single-photon
interference effects are explained in terms of second-order field
correlations \cite{MW}. The same correlation has also been referred
to as first-order in literature \cite{G-1}. We use the term
``lowest-order'' to avoid confusion.
\bibitem {Note-sp-em} In the multi-mode scenario, two pairs of
photon can also be created via spontaneous emission at a nonlinear
crystal \cite{Torren-sp-2pair}.
\bibitem {Note-ph-no-0} Average photon number is equal to zero implies
both sources are turned off. In this case, no interference occurs
simply because there is no light to detect.
\end{thebibliography}
\end{document}